\documentclass[aps,prd,twocolumn,preprintnumbers]{revtex4}

\usepackage{amsmath,graphicx,color,xcolor,amssymb}

\def\beq{\begin{equation}}
\def\eeq{\end{equation}}

\newcommand{\nc}{\newcommand}
\nc{\Tr}{\mbox{Tr}}
\nc{\hc}{\mbox{H.c.}}
\nc{\Br}{\mbox{Br}}
\nc{\ev}{\;\mathrm{eV}}
\nc{\mev}{\;\mathrm{MeV}}
\nc{\gev}{\;\mathrm{GeV}}
\nc{\tev}{\;\mathrm{TeV}}
\nc{\infinity}{\infty}
\definecolor{darkgreen}{cmyk}{0.85,0.2,1.00,0.2}

\newcommand{\svc}{\langle \sigma v^2\rangle_{\rm can}}

\begin{document}

\title{ Cannibal domination and the matter power spectrum}

\author{Adrienne L. Erickcek}
\email{erickcek@physics.unc.edu}
\affiliation{Department of Physics and Astronomy, University of North Carolina at Chapel Hill, Phillips Hall CB3255, Chapel Hill, North Carolina 27599, USA}

\author{Pranjal Ralegankar}
\email{pranjal6@illinois.edu}
\affiliation{Illinois Center for Advanced Studies of the Universe, Department of Physics, University of Illinois at Urbana-Champaign, 1110
  West Green Street, Urbana, Illinois 61801, USA}

\author{Jessie Shelton}
\email{sheltonj@illinois.edu}
\affiliation{Illinois Center for Advanced Studies of the Universe, Department of Physics, University of Illinois at Urbana-Champaign, 1110
  W. Green Street, Urbana, IL 61801, USA}

\begin{abstract}

Decoupled hidden sectors can easily and generically result in a period of cannibal domination, during which the dominant component of the Universe has an equation of state intermediate between radiation and matter due to self-heating by number-changing interactions.
We present for the first time the consequences of a cannibal-dominated era prior to big bang nucleosynthesis for structure formation on small scales.  We find that an early cannibal-dominated era imprints a characteristic peak on the dark matter power spectrum, 
with scale and amplitude directly determined by the mass, lifetime, and number-changing interaction strength of the cannibal field.
This enhancement to the small-scale matter power spectrum will generate early-forming dark matter microhalos, and we provide a detailed and transparent map between the properties 
of the cannibal species and the characteristic mass and formation time of these structures.   These relations demonstrate how the internal workings of a hidden sector leave a potentially observable imprint on the matter power spectrum even if dark matter has no direct couplings to the Standard Model. 

\end{abstract}

\maketitle

\section{Introduction}
 It is easy and generic for the early Universe to contain self-interacting hidden sectors that are thermally decoupled from the Standard Model (SM). Such decoupled hidden sectors can readily be produced during post-inflationary reheating along with the SM \cite{Hodges:1993yb,Berezhiani:1995am,Adshead:2016xxj,Adshead:2019uwj} and can significantly broaden options for dark matter (DM) models and their resulting signatures, e.g. \cite{Faraggi:2000pv,Feng:2008mu,DAgnolo:2015ujb,Berlin:2016vnh,Dror:2016rxc,Garny:2018grs,March-Russell:2020nun}.  Decoupled hidden sectors can also  give rise to nonstandard cosmological histories: when the lightest particle in the hidden sector is massive, it can easily come to temporarily dominate the energy density of the Universe, leading to departures from radiation domination prior to the onset of big bang nucleosynthesis (BBN).  If the lightest particle in the hidden sector is effectively pressureless while it dominates the Universe, then it causes an early matter-dominated era (EMDE) \cite{Zhang:2015era,Berlin:2016vnh,Berlin:2016gtr,Dror:2016rxc,Dror:2017gjq}.  An EMDE has profound implications for both the relic abundance of DM \cite{Moroi:1999zb,GKR01,Kane:2015jia,Co:2015pka,Berlin:2016vnh,Berlin:2016gtr, Blinov:2019rhb} and the abundance of sub-earth-mass DM microhalos \cite{ES11, BR14, FOW14, Erickcek:2015jza, Blinov:2019jqc}.  The observational signatures of this microhalo population are predominantly controlled by the small-scale cutoff in the matter power spectrum \cite{Erickcek:2015jza, Erickcek:2015bda, Blanco:2019eij, Delos:2019dyh}, so constraints on EMDE cosmologies cannot be disentangled from early universe microphysics.
 
In many familiar theories, from elementary scalars to confining gauge theories, the lightest hidden particle has number-changing self-interactions that enable it to maintain internal pressure support after it becomes nonrelativistic.  Such ``cannibal'' interactions \cite{Dolgov:1980uu,Dolgov:2017ujf,Carlson:1992fn} are natural properties for the lightest particle in a hidden sector with a mass gap, and meta-stable cannibal species arise frequently in models of hidden sector DM  \cite{Zhang:2015era,Pappadopulo:2016pkp,Berlin:2016vnh, Tenkanen:2016jic, Dror:2016rxc, Kopp:2016yji,Krnjaic:2017tio}.  As we show here, cannibal interactions alter the evolution of density perturbations during an early cannibal-dominated era (ECDE) compared to their evolution during an EMDE and typically determine the small-scale cutoff in the matter power spectrum.  

In contrast to previous studies of structure formation with a cannibal species \cite{Machacek:1994vg, deLaix:1995vi, Buen-Abad:2018mas} or a related toy model \cite{Das:2018ons}, the cannibal 
 here is {\em unstable}, decaying to SM particles during the Universe's first second.  Therefore, it does not contribute to the present-day dark matter abundance, which we assume to be composed of a separate species. For simplicity we consider the DM abundance to have formed prior to the  
 ECDE, as is easily realized, e.g., when DM is a thermal relic in the hidden sector.  We are thus interested in the evolution of perturbations in a three-component universe, consisting of cold decoupled DM, the SM radiation bath, and the metastable cannibal species.
 
We find that an ECDE generates a peak in the power spectrum of DM density fluctuations on scales that enter the cosmological horizon during the ECDE. 
We show that the shape, amplitude, and scale of the ECDE peak is  determined by the properties of the cannibal particle, with little sensitivity to DM particle properties.
This enhancement to the matter power spectrum will generate DM microhalos long before structure formation would occur in the absence of an ECDE, and we relate the characteristic mass and formation time of these microhalos to the mass of the cannibal field, the strength of its number-changing interactions, and its lifetime.  Our results reveal a new window into the thermal history of the pre-BBN Universe and further establish that hidden sectors can generate distinctive observational signatures even in the absence of interactions between DM and the SM.

We begin in Sec.~\ref{sec:background} with a summary of how cannibal interactions affect the homogeneous evolution of the cannibal field, including its equation of state.  In Sec.~\ref{sec:perts} we examine the evolution of density perturbations in both the cannibal field and the dark matter and show how an ECDE generates a peak in the matter power spectrum.  We discuss how the features of this peak determine the mass and formation time of the first dark matter halos in Sec.~\ref{sec:microhalos}, and we relate these quantities to the properties of the cannibal field.  We summarize our findings in Sec.~\ref{sec:summary}.  Additional technical details are provided in the Appendices.  Appendices \ref{app:bkg} and \ref{app:perts} contain the equations that govern the evolution of the homogenous fields and their density perturbations.  In Appendix \ref{app:transfer}, we describe our computation of the matter power spectrum following an ECDE.  Finally, we evaluate dark matter free streaming length in Appendix \ref{app:dmfs}.

\section{Homogeneous background evolution}
\label{sec:background} 
Cannibalism occurs when a particle has number-changing self-interactions that remain efficient even after the particle becomes non-relativistic. During cannibalism, these interactions heat the cannibal species so that its temperature $T_{\rm c}$ drops only logarithmically with scale factor once $T_{\rm c}\lesssim m/5$, where $m$ is the cannibal mass \cite{Carlson:1992fn,Pappadopulo:2016pkp}.  Meanwhile the energy density evolves as $\rho_{\rm can}\propto 1/[a^3 \ln (a/a_{\rm can})]$ \cite{Carlson:1992fn,Pappadopulo:2016pkp}, where $T_{\rm c}(a_{\rm can})\equiv m/5$.
We take as a representative cannibal model a real scalar with potential $V(\phi) = \frac{1}{2}m^2\phi^2+\frac{g}{3!}\phi^3+\frac{\lambda}{4!}\phi^4$.
This theory can also be thought of as a toy model for the lightest glueball in a pure glue hidden sector \cite{Boddy:2014yra,Boddy:2014qxa,Soni:2016gzf,Forestell:2016qhc}.
The cross-section for number-changing $3\to 2$ interactions is then \cite{Pappadopulo:2016pkp,Farina:2016llk}
\begin{align}\label{eq:alpha_can}
\svc &=\frac{25\sqrt{5}\pi^2\alpha_{\rm c}^3}{5184 m^5} + \mathcal{O}(T_{\rm c}/m),
\end{align} 
where we have defined the effective coupling $\alpha_{\rm c}$ as 
 $(4\pi\alpha_{\rm c})^3\equiv (g/m)^2[(g/m)^2+3\lambda]^2$.

\begin{figure}
\begin{center}
\includegraphics[width=0.98\linewidth]{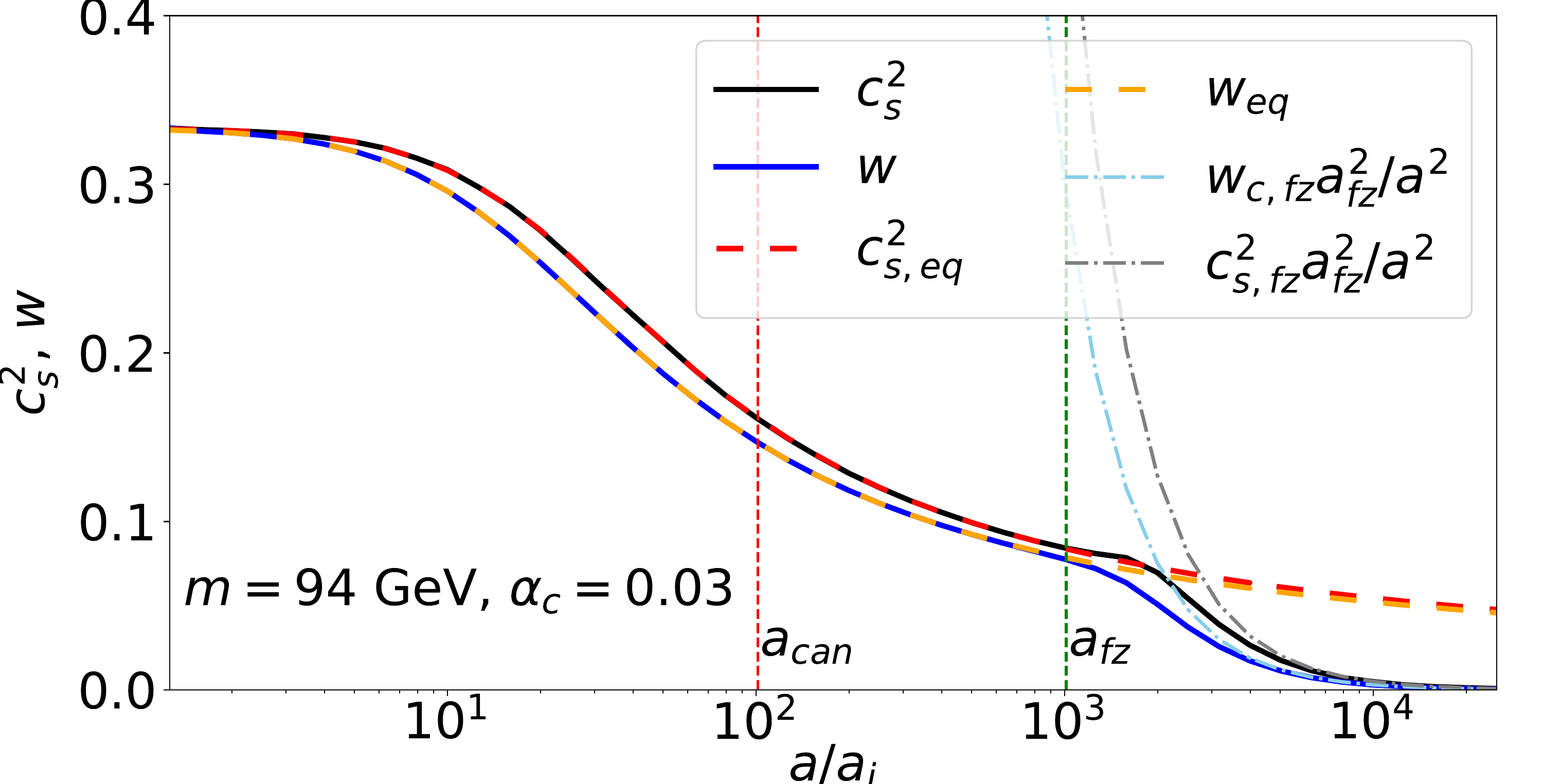}
\caption{Evolution of the cannibal sound speed $c^2_s$ and equation of state $w$ from initial cannibal temperature $T_{\rm c}(a_i)=10 m$. The red (yellow) dashed line shows the evolution of $c_s^2$ ($w$) in thermal equilibrium, the gray (cyan) dot-dashed lines show the asymptotic post-freezeout evolution, and the solid black (blue) line shows the full numerical result.
  \label{fig:csw}}
\end{center}
\end{figure}

The cannibal number density $n_\mathrm{can}$ will freeze out when the three-particle scattering rate equals the Hubble rate, which occurs at a scale factor $a_{\rm fz}$ defined by $\svc  n_{\rm can}^2(a_{\rm fz})=H(a_{\rm fz})$.
Using $n_{\rm can}(a_{\rm fz})\approx \rho_{\rm can,eq}(a_{\rm fz})/m$, where $\rho_{\rm can,eq}$ is the equilibrium cannibal density, we can estimate that the duration of cannibalism during an ECDE is $a_{\rm fz}/a_{\rm can} \sim 300\,\alpha_{\rm c}^{2/3} (\mathrm{GeV}/m)^{2/9}$.  Here, and in similar estimates below, we employ an expression for  $\rho_{\rm can,eq}$ derived in Appendix \ref{app:bkg}, 
\begin{equation}
\rho_{\rm can,eq}(a) \approx 1.5\times10^{-4} \frac{m^4  (a_{\rm can}/a)^3}{\ln(5a/a_{\rm can})}, 
\end{equation}
and we set $a_{\rm fz}/a_{\rm can}=10$ within the logarithm.
In practice, the freezeout of cannibal interactions is not instantaneous, as can be seen from Fig.~\ref{fig:csw}, where we show the evolution of the equation of state parameter $w_{\rm c} = \mathcal{P}_{\rm can}/\rho_{\rm can}$ and the sound speed $c^2_s=\partial \mathcal{P}_{\rm can}/\partial \mathcal{\rho}_{\rm can}$.
The non-negligible sound speed during cannibalism and the slow decay of the sound speed after $a_{\rm fz}$
will have important implications for perturbation growth during cannibal domination.

After cannibal freezeout, the homogeneous cannibal fluid evolves like pressureless matter until the Hubble rate becomes comparable to its decay width, $\Gamma$, at which point the cannibal decays into SM radiation.  
We define the reheat temperature, $T_{\rm rh}$, by equating the Hubble rate in the SM radiation-dominated universe to the cannibal decay rate, 
$\Gamma=\sqrt{\pi^2g_{*}(T_{\rm rh})/90} \,(T_{\rm rh}^2/M_{P})$,
where $M_P$ is the reduced Planck mass and $g_*$ is the SM radiation density divided by $[(\pi^2/30)T^4]$.
Similarly, we define the scale factor at reheating $a_{\rm rh}$ through this definition of $T_{\rm rh}$ by extrapolating adiabatic SM evolution backwards from the present day to $T_{\rm rh}$: $g_{*s}(T_{\rm rh})(a_{\rm rh}T_{\rm rh})^3 = g_{*s} (T_0) T_0^3a_0^3$,
where $T_0$ and $a_0$ are the present-day cosmic microwave background (CMB) temperature and scale factor and $g_{*S}$ is the entropy density divided by $(2\pi^2/45)T^3$ \footnote{We caution that, under these definitions, $T(a_{\rm rh})\neq T_{\rm rh}$.}.
The expansion between the onset of cannibalism and reheating then goes like $(a_{\rm rh}/a_{\rm can}) \sim 0.2 (m/T_{\rm rh})^{4/3} (10/g_*(T_{\rm rh}))^{1/3}$.

\section{Perturbation evolution}  
\label{sec:perts}
\begin{figure*}
\includegraphics[width=\textwidth]{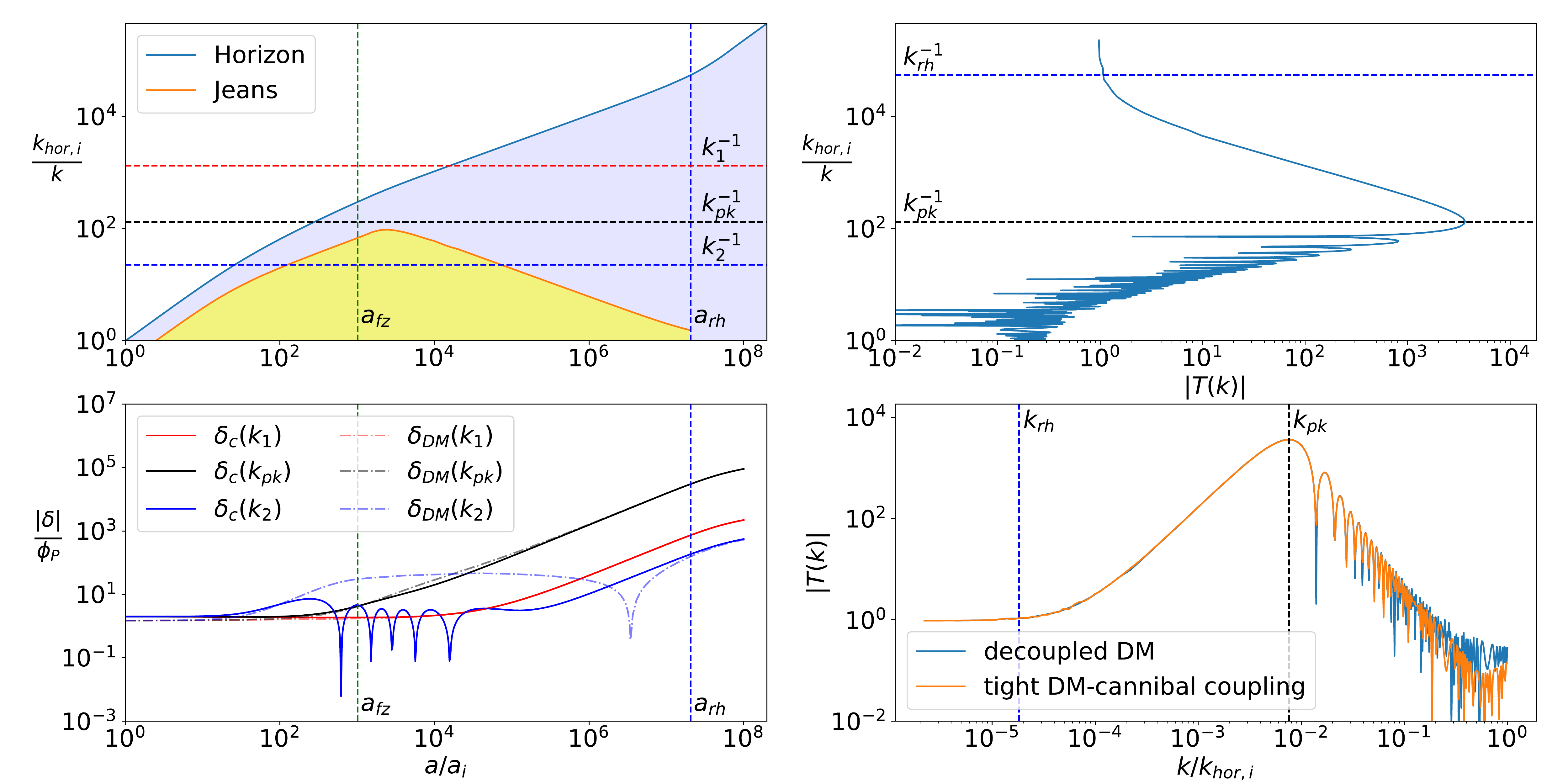}
\caption{This figure takes
$m= 1.8$~TeV, $T_{\rm rh}=10$~MeV and $\alpha_{\rm c} = 0.1$ in order to realize large separations of scales, to better highlight the physics. Results are shown for DM and cannibal fluids that only interact gravitationally, except at bottom right.  Modes that enter the horizon at $a_i$ have comoving wavenumber $k_{\rm hor,i} = a_i H(a_i)$.
\textbf{Top left}: evolution of the comoving horizon scale (solid blue) and  cannibal Jeans length (solid orange) relative to $k_{\rm hor,i}^{-1}$. Horizontal dashed lines mark the comoving wavenumbers corresponding to the perturbations $\delta_{c}$ and $\delta_{\rm DM}$ shown in the bottom left panel. 
\textbf{Bottom left}: evolution of $\delta_{\rm c}$ (solid) and $\delta_{\rm DM}$ (dash-dotted) for three selected comoving wavenumbers, relative to primordial metric perturbation amplitude $\phi_p$.
\textbf{Bottom right}: absolute value of the transfer function. The vertical blue dashed line marks the mode that enters the horizon at $a_{\rm rh}$, while the vertical black dashed line marks the mode corresponding to $k_{pk}^{-1}=1.4 k_J^{-1}(2a_{fz})$. The blue curve shows results for DM coupled only gravitationally to the cannibal, while the orange curve shows results for DM tightly coupled to the cannibal through reheating.
\textbf{Top right}: same as bottom right panel, but with axes inverted for comparison with top left panel. 
\label{fig:cann_dom_horz_peak}}
\end{figure*}

We work in conformal Newtonian gauge and follow the conventions of Ref.~\cite{Ma:1995ey}, considering the evolution of $\delta_i \equiv [\rho_i(t,x)-\bar \rho_i(t)]/\bar \rho_i(t)$, the local density fluctuation in fluid $i$, and $\theta_i\equiv\partial_{j}v_i^j$, which is the comoving divergence of the physical fluid velocity, $v_i^j=a \, dx^j/dt$.  
The resulting suite of perturbation equations is given in Appendix~\ref{app:perts}.
We set initial conditions when the cannibal is still relativistic, defining $a_i$ through $T_{\rm c}(a_i)= 10 m$, and take the cannibal to dominate the energy density of the Universe at $a_i$.  For simplicity we take the DM to be already nonrelativistic with fixed comoving density at $a_i$. We impose adiabatic initial conditions for all perturbations, as expected in the minimal cosmology where both the SM and the dark sector are populated through the decays of a single inflaton field.
We first consider the case where the cannibal fluid interacts with DM only gravitationally, and subsequently comment on the case where DM is tightly coupled to the cannibal fluid through reheating.
Numerically solving the perturbation equations through the ECDE 
until well into radiation domination yields the perturbation evolution shown in Fig.~\ref{fig:cann_dom_horz_peak}. 
%
%

The pressure generated by cannibalization, which can linger substantially after $a_{\rm fz}$, supports oscillations in the cannibal fluid.  We show in Appendix~\ref{app:perts} that the evolution of subhorizon cannibal density perturbations during the ECDE is approximately given by
\beq
\label{eq:delta_can_subhz}
\delta_{\rm c}''+\Big[\frac{(a^2 {H})'}{a^2 {H}}+\frac{(1-3w_{\rm c})}{a}\Big]\delta_{\rm c}' 
 +\frac{3(1+w_{\rm c})}{2a^2 }\Big[\frac{ {k}^2}{k_J^2}-1\Big]\delta_{\rm c}=0,
\eeq
where primes indicate derivatives with respect to $a$.  Here we have identified a comoving scale $k_J$,
\begin{align}\label{eq:jeans}
    k_J(a)=\sqrt{\frac{3}{2}(1+w_{\rm c})}\ \frac{aH}{c_s},
\end{align}
which separates oscillatory modes from growing modes.  The corresponding Jeans horizon $1/k_J$ is shown in orange in Fig.~\ref{fig:cann_dom_horz_peak}, while the comoving horizon scale $1/(aH)$ is shown in blue.
The Jeans horizon grows while the cannibal is in equilibrium and then decreases for $a \gtrsim 2a_{\rm fz}$ following the post-freezeout decline of $w_{\rm c}$ and $c_s^2$ seen in Fig.~\ref{fig:csw}. For $a\gg a_{\rm fz}$, $c_s^2\propto 1/a^2$ while the comoving horizon grows as $(aH)^{-1}\propto a^{1/2}$, yielding $k_J^{-1}\propto a^{-1/2}$. 
Cannibal perturbations will oscillate while $k>k_J (a)$, as shown, for instance, by the mode with wavenumber $k_2$ in Fig.~\ref{fig:cann_dom_horz_peak}: the oscillations seen in the bottom left panel start when the mode enters the Jeans horizon on the top panel, and cease when the mode exits. 
Since the Jeans horizon decreases slowly after freezeout, cannibal density perturbations may continue to oscillate for many  $e$-folds of expansion after cannibal freezeout. Once the Jeans horizon becomes smaller than the scale of the perturbation, i.e. when $k<k_J(a)$, the mode will stop oscillating and begin the linear growth characteristic of matter density perturbations in a matter-dominated era.

During the ECDE, the cannibal fluid dominates the gravitational potential.  As long as a given cannibal mode is oscillating, the DM perturbation sees no net gravitational potential and does not grow, as seen in the bottom left panel of Fig.~\ref{fig:cann_dom_horz_peak}. 
Once the cannibal perturbations exit the Jeans horizon and begin to grow, the DM starts to fall into the gravitational potential wells sourced by the cannibal, asymptoting to $\delta_{\rm DM}=\delta_{\rm c}$ during this matter-dominated phase.  After reheating, $\delta_{\rm DM}$ then grows logarithmically as expected in a radiation-dominated universe.  Cannibal modes that enter the horizon after $\sim\!\! 2a_{\rm fz}$, on the other hand, are well approximated as pressureless matter, and hence give rise to linear growth in cannibal as well as DM perturbations prior to reheating.  

The resulting growth experienced by a given DM mode in an ECDE, $\delta_{\rm DM}(k,a)$, compared to the prediction from uninterrupted radiation domination, $\delta_{\rm DM,s}(k,a)$, is quantified by the  transfer function \mbox{$T(k) \equiv \delta_{\rm DM}(k,a)/\delta_{\rm DM,s}(k,a)$} evaluated after matter-radiation equality, which is derived in Appendix~\ref{app:transfer} and shown in the right panels of Fig.~\ref{fig:cann_dom_horz_peak}. The DM mode that experiences the greatest growth during the ECDE, $k_{\rm pk} \sim k_{J,\textrm{min}}$, is the smallest-scale mode that never enters the cannibal Jeans horizon, 
as this mode sees the maximum amount of linear growth. Modes that enter later spend fewer $e$-folds of expansion inside the horizon, while modes that enter earlier remain inside the Jeans horizon for some time after $2a_{\rm fz}$, and thus do not begin growing until some $a>2a_{\rm fz}$.   Empirically we find $k_{\rm pk}^{-1} \approx 1.4 k_J^{-1}(2a_{\rm fz})$: this estimate is accurate to within 30\% for $a_\mathrm{rh}=5a_\mathrm{fz}$ and reaches 3\% accuracy for larger $a_\mathrm{rh}/a_\mathrm{fz}$.

Depending on the phase of the cannibal oscillation when a given cannibal mode exits the Jeans horizon, overdense regions at this time may correspond to regions that were initially either overdense or underdense.  If DM and cannibals interact only gravitationally, 
DM particles will remain in regions that were initially overdense until the cannibal stops oscillating, at which point they will fall into the gravitational wells sourced by regions with high cannibal density.  The relative phase differences between DM and cannibal perturbations at the time of Jeans-horizon exit are responsible for the oscillations exhibited in the transfer function for $k > k_{\rm pk}$, as illustrated by the evolution of the mode $k_2$ shown in Fig.~\ref{fig:cann_dom_horz_peak}.
For this mode, cannibal oscillations cease when initially underdense regions are overdense, and so the sign of $\delta_{\rm DM}$ changes as the DM particles fall into these regions.  In the alternative scenario where the cannibal and DM are tightly coupled through reheating, DM perturbations will instead oscillate in tandem with the cannibals throughout the ECDE; the resulting oscillatory cutoff in the transfer function is the same, as seen in the bottom right panel of Fig.~\ref{fig:cann_dom_horz_peak}.

Using $k_{\rm pk}^{-1} = 1.4 k_J^{-1}(2a_{\rm fz})$ and taking \mbox{$1+w_{\rm c}\approx 1$}, $c_s^2\approx T_c/m $ near $a_{\rm fz}$, the maximum perturbation growth occurs for $k_{\rm pk} \sim 0.3\sqrt{a_i/a_{\rm fz}} k_{\rm hor,i}$, provided that $a_{\rm fz}\ll a_{\rm rh}$.  In terms of the cannibal properties, we then have
\beq
\label{eq:kpk}
k_{\rm pk} \sim 40 \mathrm{pc}^ {-1}  \left(\frac{\alpha_{\rm c}}{0.02}\right)^{-1/3} \left(\frac{T_{\rm rh}}{10 \mev}\right)^{1/3} \left(\frac{m}{\tev}\right)^{7/9}.
\eeq
The corresponding maximum of the transfer function is proportional to the maximum duration of linear growth; we find $T(k_{\rm pk}) \sim (1/5) a_{\rm rh}/a_{\rm fz}$, or
\beq
\label{eq:Tkpk}
T(k_{\rm pk})\sim 5\times 10^{3}\left(\frac{\alpha_{\rm c}}{0.02}\right)^{-2/3}\Big(\frac{m}{\tev}\Big)^{14/9}\left(\frac{T_{\rm rh}}{\rm 10\, MeV}\right)^{-4/3}.
\eeq

\section{Implications for microhalo formation}  
\label{sec:microhalos}
Early enhancements to perturbation growth lead to the early formation of sub-earth-mass microhalos, with well over half of all DM potentially residing in highly concentrated microhalos at high redshift ($z\gtrsim20$) if DM is sufficiently cold \cite{ES11, Erickcek:2015jza}.  Moreover, their high densities imply that most of these microhalos survive within galaxies: they may lose a significant fraction of their mass, but their dense centers remain intact \cite{Delos:2019lik, Delos:2019tsl, Delos:2019dyh}. 

The microhalo mass corresponding to $k_{\rm pk}$ can be estimated as $M_{\rm pk}\equiv\frac{4\pi}{3}(k_{\rm pk})^{-3}\bar\rho_m$
where $\bar\rho_m$ is the mean present-day matter density, giving
\begin{align}
\frac{M_{\rm pk}}{M_{\odot}}&\sim 3\times 10^{-12}\left(\frac{\alpha_{\rm c}}{0.02}\right)\left(\frac{\rm 10 \,MeV}{T_{\rm rh}}\right)\left(\frac{\tev}{m}\right)^{7/3}.
\end{align}
Since $T(k)$ exhibits a relatively narrow peak due to the cutoff arising from cannibal acoustic oscillations, the first microhalos that form have masses very close to $M_{\rm pk}$. 
 
The collapse of overdense regions into gravitationally bound microhalos occurs when $\delta_{\rm DM}$ exceeds order unity.  Therefore, the maximum value of the transfer function $T(k_{\rm pk})$ determines when the first microhalos form.  In the absence of an ECDE, earth-mass microhalos form around a redshift of 60 \cite{Diemand:2005vz}, 
with smaller halos forming only slightly earlier.  Since perturbations on scales smaller than the baryon Jeans length grow as $\delta_{\rm DM} \propto (1+z)^{-0.9}$ during matter domination \cite{Hu:1995en, Bertschinger06}, the formation redshift of the first microhalos following an ECDE will be roughly a factor of $[T(k_{\rm pk})]^{1.11}$ higher than 
in the absence of an ECDE for $T(k_{\rm pk}) \lesssim 30$.  Halos that form during matter domination have characteristic central densities that are proportional to the matter density at their formation \cite{delos2019predicting}, so an ECDE generates microhalos that are $[T(k_{\rm pk})]^{3.33}$ times denser than standard microhalos.
If $T(k_{\rm pk})\gtrsim 30$, the first microhalos form during radiation domination, yielding even higher central densities \cite{Blanco:2019eij}.

\begin{figure}
\begin{center}
\includegraphics[width=\linewidth]{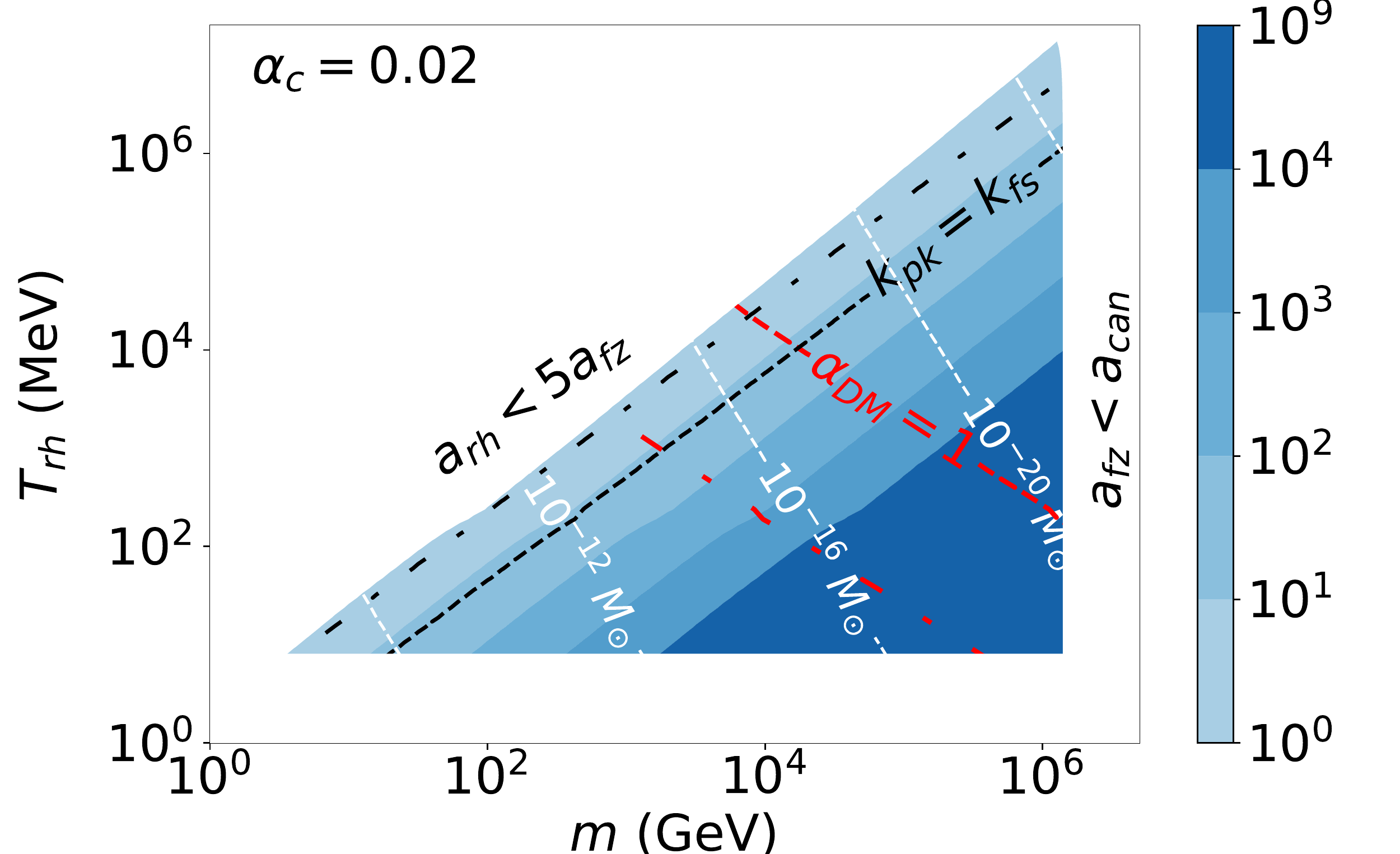}
\caption{Colored contours of $T(k_{\rm pk})$ as a function of $m$ and $T_{\rm rh}$ for fixed $\alpha_{\rm c}=0.02$. White contours show the microhalo mass $M_{\rm pk}$ corresponding to $k_{\rm pk}$.  Above the red lines thermal-relic DM becomes nonperturbative for fixed $m_{\rm DM}/m=10$ (dashed) and $100$ (dot-dashed). Above the black lines, DM free streaming determines the small-scale cutoff for fixed $m_{\rm DM}/m=10$ (dashed) and $100$ (dot-dashed) if DM kinetically decouples from the cannibal after cannibal freezeout.
\label{fig:param_space}}
\end{center}
\end{figure}

The expressions for $k_{\rm pk}$ and $T(k_{\rm pk})$ in Eqs.~\eqref{eq:kpk} and~\eqref{eq:Tkpk} thus allow microhalo properties to be directly related to cannibal particle properties. 
We show $T(k_{\rm pk})$ as a function of cannibal mass and reheat temperature in Fig.~\ref{fig:param_space}, together with contours indicating the corresponding microhalo masses $M_{\rm pk}$.  In this figure we approximate $k_{\rm pk}^{-1} = 1.4 k_J^{-1}(2a_{\rm fz})$ and subsequently determine $T(k_{\rm pk})$ by numerically solving the relevant perturbation equations for the mode $k_{\rm pk}$.

The region of cannibal properties that gives rise to enhanced microhalo formation is bounded. 
First, BBN and the CMB require $T_{\rm rh}>8.1 $ MeV \cite{deSalas:2015glj,Hasegawa:2019jsa}.
Second, attaining an epoch of cannibal domination requires 
$a_{\rm can}<a_{\rm fz}$, which 
yields an upper bound on the cannibal mass in terms of $\alpha_{\rm c}$: $m\lesssim 10^{11}\alpha_{\rm c}^3~ {\rm GeV}$. 
Meanwhile, the hierarchy $a_{\rm rh}/a_{\rm fz}$ controls the duration of the EMDE that follows cannibal freezeout and thus the value $T(k_{\rm pk})$. 
Fig.~\ref{fig:param_space} shows results for $a_{\rm rh}>5 a_{\rm fz}$,  or $m/\mathrm{GeV} \gtrsim 20\alpha_{\rm c}^{3/7}(T_{\rm rh}/{\rm 10\, MeV})^{6/7} (g_*(T_{\rm rh})/10)^{3/14}$ . In this regime neglecting cannibal decays around $a_{\rm fz}$, as assumed in our numerical work, is a good approximation.  Models with $a_{\rm rh} < 5a_{\rm fz}$ are physically possible, but, as Fig.~\ref{fig:param_space} makes clear, do not exhibit interesting enhancements to small-scale structure.  
When $T(k_{\rm pk})\gtrsim 10^4$, density perturbations collapse during the ECDE.  The decay of the cannibal into radiation destroys these structures, and the subsequent free streaming of the released DM particles suppresses later microhalo formation \cite{Blanco:2019eij}.  DM free streaming \cite{Green_2005, Loeb_2005, Bertschinger06, Piattella:2013cma, Hogan:2000bv} can also be relevant for $T(k_{\rm pk}) \lesssim 10$, as shown in Fig.~\ref{fig:param_space} and discussed in Appendix~\ref{app:dmfs}.

Additionally, not all of this parameter space is consistent with DM production via thermal freezeout.  If DM self-annihilates with cross-section $\langle\sigma v\rangle = \pi \alpha_D^2/m_{\rm DM}^2$, higher DM masses demand larger couplings to obtain the observed DM abundance.  Entropy production during the ECDE dilutes the DM abundance,
but requiring $\alpha_D<1$ still establishes an upper bound on $m_{\rm DM}$.  Furthermore, our calculations assume that DM is already nonrelativistic at $T_i = 10 m$, and thus that $m_{\rm DM}\gg m$.  Since higher values of $T_{\rm rh}$ require higher cannibal masses to ensure cannibal freezeout prior to reheating, requiring $\alpha_D<1$ establishes an upper bound on $T_{\rm rh}$, as depicted by the black dashed and dot-dashed lines in Fig.~\ref{fig:param_space}.  In the region above these lines, other production mechanisms for DM must be invoked.   

\section{Summary}
\label{sec:summary}
An early cannibal-dominated era (ECDE) is a generic feature of a decoupled hidden sector; it occurs whenever the lightest hidden-sector particle comes to dominate the energy density of the Universe while it is nonrelativistic and undergoing number-changing interactions.  We present for the first time the consequences of an ECDE for the evolution of density perturbations and the formation of DM halos.

Perturbation growth is suppressed for modes that enter the cannibal Jeans horizon, providing a novel and distinctive small-scale cutoff in the matter power spectrum.  Modes that enter the horizon after cannibal freezeout but before cannibal decay, on the other hand, experience the same growth as modes that enter the horizon during an EMDE.  The location and amplitude of the resulting peak in the matter power spectrum can be computed from the cannibal's mass, lifetime, and self-interaction cross-section.  Moreover, as long as the DM relic abundance is in place prior to the ECDE, the DM mass and interactions with the cannibal field have no significant effects on the matter power spectrum throughout most of the relevant parameter space.  We are thus able to provide a precise and transparent map between cannibal particle properties and the cutoff scale. In contrast, prior analyses of the matter power spectrum following an EMDE have employed rough estimates of the cutoff scale using either the DM free-streaming length \cite{ES11,FOW14,Erickcek:2015jza, Dror:2017gjq,Erickcek:2015bda} or the mass of the particle that induces the EMDE \cite{Zhang:2015era,Blanco:2019eij}. 

The cutoff scale determines both the location and the amplitude of the peak in the matter power spectrum; the peak scale then sets the mass of the first generation of microhalos while the peak amplitude sets their abundance and formation time \cite{Erickcek:2015jza}.  Since early-forming microhalos are denser than later-forming microhalos \cite{delos2019predicting}, higher-amplitude peaks generate microhalos with stronger lensing and annihilation signatures, up to the point that halos form during the ECDE itself.  We are therefore able to identify which regions of cannibal parameter space predict interesting signals for observational probes of the microhalo population such as pulsar timing arrays \cite{dror2019pta, Ramani:2020hdo}, stellar microlensing within galaxy clusters \cite{Dai:2019lud}, and the contribution to the isotropic gamma-ray background from dark matter annihilations \cite{Blanco:2019eij, Delos:2019dyh}.  This work shows that the particle physics of dark sectors, beyond dark matter itself, can control the small-scale structure of DM in our universe.
\medskip

\acknowledgments{ALE is supported in part by NSF CAREER grant PHY-1752752. The work of JS and PR is supported in part by DOE CAREER grant DE-SC0017840.  Portions of this work were performed at the Aspen Center for Physics, which is supported by National Science Foundation grant PHY-1607611, and at the Kavli Institute for Theoretical Physics, which is supported in part by the National Science Foundation under Grant No. NSF PHY-1748958.}

\appendix
\section{Homogeneous Background Equations}
\label{app:bkg}

The Boltzmann equations that describe the evolution of homogeneous radiation and cannibal densities are
\begin{align}
	\frac{d\rho_{\rm can}}{dt}&+3H\rho_{\rm  can}(1 +w_c(a))=-\Gamma m n_{\rm  can}\label{eq:num_can_density}\\
	\frac{d\rho_r}{dt}&+4H\rho_r=\Gamma m n_{\rm can}\label{eq:num_r_density}\\
	\frac{dn_{\rm  can}}{dt}&+3Hn_{\rm  can}=\svc  n_{\rm can}^2(n_{\rm  eq,can}-n_{\rm  can})-\Gamma n_{\rm can}\label{eq:num_can_ndensity},
\end{align}
where the Hubble rate is given by $H=\sqrt{\rho_{\rm can}+\rho_r+\rho_{\rm DM}}/(\sqrt{3}M_{P})$. The DM density evolves adiabatically ($\rho_{\rm DM}\propto 1/a^3$) throughout our analysis.  Note that Eq.~\eqref{eq:num_can_ndensity} implies that $n_{\rm can}$ decreases after the cannibal becomes nonrelativistic, while Eq.~\eqref{eq:num_can_density} implies that the cannibal's energy density is conserved if decays are neglected.  

We assume that two-to-two cannibal interactions are fast enough to maintain internal kinetic equilibrium throughout our analysis. The cannibal fluid can thus be fully described in terms of its chemical potential $\mu$ and temperature $T_c$:
\begin{align}\label{eq:can_density_eq}
	\rho_{\rm can}&=\int \frac{d^3p}{(2\pi)^3}E f\left(\frac{E-\mu}{T_c}\right); \\
	w_c(a)&=\frac{\int \frac{d^3p}{(2\pi)^3}\frac{p^2}{3E} f\left(\frac{E-\mu}{T_c}\right)}{\int \frac{d^3p}{(2\pi)^3}E f\left(\frac{E-\mu}{T_c}\right)}; \\ 
	n_{\rm can}&=\int \frac{d^3p}{(2\pi)^3}f\left(\frac{E-\mu}{T_c}\right),
\end{align}
where $f(x)=(e^{x}-1)^{-1}$ is the Bose-Einstein distribution. For $a\ll a_{\rm fz}$, the cannibal is in chemical equilibrium, i.e. $\mu=0$, and Eq.~\eqref{eq:num_can_density} is sufficient to describe the equilibrium cannibal density.

We consider scenarios with $a_{\rm rh}>5a_{\rm fz}$.  In this case cannibal decays are negligible for $a\ll a_{\rm fz}$ and the conservation of the comoving cannibal entropy, $S_c$, enables a useful parametrization of the evolution of the equilibrium cannibal fluid after it becomes non-relativistic \cite{Carlson:1992fn,Buen-Abad:2018mas}, 
\begin{align}
	T_{c,\textrm{eq}}&\approx \frac{m}{3\ln\left(\frac{mS_{c}^{-1/3}}{\sqrt{2\pi}}a\right)}; \\ 
	\rho_{\rm can,eq}&\approx \frac{mS_{c}}{3a^3\ln\left(\frac{mS_{c}^{-1/3}}{\sqrt{2\pi}}a\right)}\\
	&\approx \frac{150 m^4}{\ln \left(\frac{a}{20 a_i}\right)} \, \left(\frac{a_i}{a}\right)^3.
	\label{eq:approxevol}
\end{align}
Here in the last relation we have used that $S_c$ can be evaluated at $a_i$, when the cannibal is still relativistic:  $S_c=2\pi^2(a_iT_c(a_i))^3/45=2\pi^2(10ma_i)^3/45$. Equation~\eqref{eq:approxevol} is a good approximation for cannibal evolution 
after $T_c$ falls below $m/5$, which occurs when $a = a_{\rm can}\approx100a_i$.

\section{Perturbation Equations}
\label{app:perts}

The three components of our model (the cannibal, the dark matter, and SM relativistic particles) are approximated as perfect fluids with no anisotropic stresses.  
Perfect fluids can be described with two perturbation fields: the density perturbation $\delta_i\equiv [\rho_i(t,x)-\bar\rho_i(t)]/\bar\rho_i(t)$ relative to the homogeneous background density $\bar \rho_i$, and the comoving divergence of the physical fluid velocity, $\theta_i=\partial_{j}v_i^j$ where $v^j_i=a\,dx^j_i/dt$.  We work in conformal Newtonian gauge and use the perturbed spacetime metric
\begin{align}
ds^2=-\left[1+2\psi(\vec{x},t)\right]dt^2+a^2(t)\left[1-2\phi(\vec{x},t)\right]d\vec{x}^2.
\end{align}
In the absence of anisotropic stress, $\psi=\phi$.

To obtain evolution equations for the perturbations $\delta$ and $\theta$, we perturb the divergence of the energy momentum tensor $\nabla_\mu T^{\mu\nu} = Q^\nu$ for each fluid, as described in Ref.~\cite{ES11}.  In the absence of cannibal decays, $Q^\nu = 0$ for all fluids, and our equations recover those derived in Refs.~\cite{Ma:1995ey,Buen-Abad:2018mas}.   Due to the time dilation experienced by moving particles, the (homogeneous) collision operator describing the energy lost from the cannibal fluid due to decays can be written as $\Gamma n_{\rm can} m$, where $n_{\rm can}$ is the cannibal number density and $\Gamma$ its width.  Here we make the approximation that $\Gamma n_{\rm can} m \approx \Gamma \rho_{\rm can}$ to facilitate obtaining a covariant formulation of $Q^\nu$ in terms of the energy-momentum tensor.  This approximation corresponds to neglecting terms in the perturbation equations that are proportional to $(\Gamma/H) w_{\rm c}$ and $(\Gamma/H) c_s^2$, but since we only consider scenarios in which reheating occurs well after cannibal interactions freeze out ($a_\mathrm{rh} > 5 a_\mathrm{fz}$), these terms have no significant impact on the peak in the matter power spectrum that follows an ECDE.  With this approximation, the evolution of perturbations in our three-fluid system is determined by the following suite of equations:
\begin{widetext}
\begin{align}
    \delta_{\rm c}'(a)&=-(1+w_{\rm c})\left(\frac{ {\theta_{\rm c}}}{a^2 {H}}-3\phi'\right)-\frac{3}{a}(c_s^2-w_{\rm c})\delta_{\rm c}-\frac{ {\Gamma}}{a {H}}\phi,\label{eq:delta_can_eq}\\
     {\theta_{\rm c}}'(a)&=-\frac{1}{a}(1-3w_{\rm c})\theta_{\rm c}-\frac{w_{\rm c}'}{1+w_{\rm c}} {\theta_{\rm c}}+\frac{c_s^2}{1+w_{\rm c}}\frac{ {k}^2}{a^2 {H}}\delta_{\rm c}+\frac{ {k}^2}{a^2 {H}}\phi\label{eq:theta_can_eq}\\
    \delta_{\rm DM}'(a)&=-\frac{ {\theta}_{\rm DM}}{a^2 {H}}+3\phi',\label{eq:delta_dm_eq}\\
     {\theta}_{\rm DM}'(a)&=-\frac{1}{a} {\theta}_{\rm DM}+\frac{ {k}^2}{a^2 {H}}\phi + \frac{\gamma}{aH}\left(\theta_{\rm c} - \theta_{\rm DM}\right)\label{eq:theta_dm_eq}\\
    \delta_r'(a)&=-\frac{4}{3}\frac{ {\theta}_r}{a^2 {H}}+4\phi'+\frac{ {\Gamma}\rho_{\rm can}}{a {H}\rho_r}(\delta_{\rm c}-\delta_r+\phi)\label{eq:delta_r_eq}\\
     {\theta}_r'(a)&=\frac{1}{4}\frac{ {k}^2}{a^2 {H}}\delta_r+\frac{ {k}^2}{a^2 {H}}\phi+\frac{ {\Gamma}\rho_{\rm can}}{a {H}\rho_r}\Big(\frac{3}{4}\theta_{\rm c}-\theta_r\Big)\label{eq:theta_r_eq}\\
          {k}^2\phi+3(a {H})^2\Big(a\phi'+\phi\Big)&=-\frac{1}{2}a^2\frac{1}{M_{P}^2}\left(\rho_{\rm can}\delta_{\rm c}+\rho_r\delta_r+\rho_{\rm DM}\delta_{\rm DM}\right)\label{eq:metric_eq}.
\end{align}
\end{widetext}
Here primes indicate derivatives with respect to $a$ and the subscripts $\mathrm{c},r$ and $\mathrm{DM}$ correspond to perturbations of the cannibal, SM radiation and DM fluids respectively.  We use Fourier-transformed variables with $k$ being the comoving Fourier mode.

Equation~(\ref{eq:theta_dm_eq}) includes scattering interactions between the DM and cannibal particles, with a scattering rate given by $\gamma = n_{\rm can} \sigma_{\rm DMc} v_{\rm c}$, where $\sigma_{\rm DMc}$ is the cross section for DM-cannibal scattering.  We neglect the DM-cannibal scattering term in Eq.~(\ref{eq:theta_r_eq}) because it is suppressed by a factor of $\rho_{\rm DM}/\rho_{\rm can}$, which is very small during the ECDE.   DM-cannibal interactions are generically $s$-wave, so in the non-relativistic limit we expect $\langle \sigma_{\rm DMc}v\rangle \propto \sqrt{T_{\rm c}}$, implying that $\gamma \propto \rho_{\rm can} \sqrt{T_{\rm c}/m}$.  The specific value of $\sigma_{\rm DMc}$ is model dependent, and in the bottom right panel of Fig.~\ref{fig:csw} we show the transfer function for two limiting cases: one with $\gamma \ll H$ while $T_{\rm c} \lesssim 10m$ and one with $\gamma \gg H$ throughout the ECDE.  While $\gamma \gg H$, DM-cannibal interactions force $\delta_{\rm DM}\approx \delta_{\rm c}$.  However, for perturbation modes that exit the cannibal Jeans horizon prior to reheating, the DM particles fall into the gravitational wells created by the cannibal particles and $\delta_{\rm DM}(a_{\rm rh}) \approx \delta_{\rm c}(a_{\rm rh})$ even if $\gamma = 0$.  Therefore, scatterings between DM and cannibal particles have no impact on the peak in the power spectrum generated by an ECDE.

We will focus on modes entering the horizon after $a_i$. Note that at $a_i$ the cannibal is still relativistic, because $a_i$ is defined by the condition that $T(a_i)=10m$. We consider the perturbations to be adiabatic, which implies that the initial conditions for super-horizon modes at $a_i$ are 
\begin{align}\label{eq:initial_condition}
\delta_r&=\frac{4}{3}\delta_{\rm DM}=\delta_{\rm c}=-2\phi_p; \\ \theta_r&=\theta_{\rm DM}=\theta_{\rm c}=\frac{1}{2}\frac{k^2}{aH}\phi_p,
\end{align}
where $\phi_p$ is the primordial metric perturbation. This adiabatic initial condition for both hidden sector and SM perturbations is naturally obtained in the minimal scenario where both sectors originate from the decays of a single inflaton field.

The approximate equation describing cannibal evolution deep inside the horizon during the ECDE, Eq.~(2) of the main paper, is obtained by using Eq.~\eqref{eq:theta_can_eq} to eliminate $\theta_{\rm c}$ in Eq.~\eqref{eq:delta_can_eq}, under the following approximations: (i) neglecting derivatives of $w_{\rm c}$ and $c_s^2$, as both quantities are slowly varying until they become negligibly small; (ii) neglecting terms proportional to $c_s^2-w_{\rm c}$; (iii) dropping terms proportional to $\Gamma/H$; and (iv) neglecting $\phi'$ in Eq.~\eqref{eq:delta_can_eq}, since deep inside the horizon that term is negligible in comparison to $\theta_{\rm c}/(a^2H)$.  Then, using Eq.~\eqref{eq:metric_eq} to eliminate  $\phi$ in the limit  $k\gg aH$, we obtain Eq.~(2).  An analogous approximate equation for DM perturbations can be derived using the same approximations.  Setting $\gamma =0$, we find
\begin{align}
\label{eq:delta_DM_subhz}
\delta_{\rm DM}''(a)&+\Big[\frac{(a^2H)'}{a^2H}+\frac{1}{a}\Big]\delta_{\rm DM}'=\frac{3}{2a^2}\frac{\rho_{\rm can}}{\rho} \delta_{\rm c}.
\end{align}
Here the rhs arises from the metric perturbation term in Eq.~\eqref{eq:theta_dm_eq}.  When $k > k_J$, $\delta_{\rm c}$ is oscillating (and thus sources an oscillatory potential $\phi$) and its time-averaged contribution to the above equation is approximately zero. Making the approximation that the cannibal density dilutes as $1/a^3$, the coefficient of $\delta_{\rm DM}'$ in Eq.~\eqref{eq:delta_DM_subhz} becomes $3/(2a)$. It follows that there are two solutions for dark matter perturbations while $\delta_{\rm c}$ is oscillating:  $\delta_{\rm DM}\propto a^{-1/2}$ and $\delta_{\rm DM}\propto a^0$. Consequently, $\delta_{\rm DM}$ remains constant while $\delta_{\rm c}$ is oscillating, as seen in Fig.~\ref{fig:cann_dom_horz_peak}.

The SM radiation is well-described as a perfect fluid throughout the ECDE because the SM particles form a tightly coupled plasma.  However, the perfect fluid approximation can break down for both the cannibal and DM particles, which both deviate from perfect-fluid behavior after they kinetically decouple.  The impact of the DM particles' random thermal motions will be considered in Appendix~\ref{app:dmfs}.  The cannibal particles generally remain kinetically coupled long after the freezeout of number-changing interactions, but may lose internal kinetic equilibrium prior to their decay in some models.  We have verified that even in these cases, the thermal motion and dispersion of the cannibal particles cannot affect the location and amplitude of the peak in the ECDE transfer function.  The relevant dispersion scales never significantly exceed the cannibal Jeans horizon at the time of kinetic decoupling, which is always much smaller than the peak scale since the Jeans horizon shrinks after cannibals freeze out.

\section{The ECDE Transfer Function}
\label{app:transfer}
We are ultimately interested in the modification to the late-time linear matter power spectrum, encapsulated in the transfer function $T(k)$. To compute this quantity, we first recall how DM perturbations that enter the horizon during radiation domination evolve as the Universe transitions from radiation domination to matter domination. After entering the horizon at some scale factor $a_{\rm hor}$, DM density perturbations initially evolve as 
\begin{align}
\label{eq:log_growth}
    \delta_{\rm DM}(a \ll a_\mathrm{eq})=-A_{\rm s}\phi_p(k)\ln\left(\frac{B_{\rm s}a}{a_{\rm hor}}\right)
\end{align}
where $A_{\rm s}=9.11$ and $B_{\rm s}=0.594$ are numerical fitting factors \cite{Hu:1995en}, and $a_{\rm eq}$ is the scale factor at matter-radiation equality. 
The logarithmic growth of $\delta_{\rm DM}$ continues until matter-radiation equality, after which $\delta_{\rm DM}$ grows linearly with the scale factor.  
The evolution of $\delta_{\rm DM}$ after matter-radiation equality can be determined by solving the Meszaros equation, 
which is a second-order homogeneous differential equation for $\delta_{\rm DM}(a)$ that is applicable while $\rho_r \delta_r \ll \rho_{\rm DM}\delta_{\rm DM}$, and then using Eq.~(\ref{eq:log_growth}) to set initial conditions for $\delta_{\rm DM}(a \ll a_{\rm eq})$ and $\delta_{\rm DM}^\prime(a \ll a_{\rm eq})$ \cite{Hu:1995en}.  After matter-radiation equality, $\delta_{\rm DM}$ is then dominated by the growing solution to the Meszaros equation:
\begin{align}
\label{eq:Meszaros}
\delta_{\rm DM}(a\gg a_{\mathrm{eq}})=-\frac{3A_{\rm s}\phi_p(k)}{2}\ln\left(\frac{4B_{\rm s}e^{-3}a_{\rm eq}}{a_{\rm hor}}\right)D_1(a/a_{\rm eq}) 
\end{align}
where $D_1(y)=2/3+y$.  

In a cosmology with an ECDE, modes that entered the horizon prior to $T_{\rm rh}$ will grow logarithmically during the period of radiation domination that follows after $T_{\rm rh}$.  The evolution of these modes while $\rho_r \delta_r \gg \rho_{\rm DM}\delta_{\rm DM}$ can also be described by Eq.~\eqref{eq:log_growth} with the numerical factors $A_{\rm s}$ and $B_{\rm s}$ replaced by $k$-dependent values $A(k)$ and $B(k)$ that encode the evolution of $\delta_{\rm DM}$ prior to reheating.  Similarly, after matter-radiation equality the evolution of $\delta_{\rm DM}$ can be described using Eq.~\eqref{eq:Meszaros} with $A_{\rm s}$ and $B_{\rm s}$ replaced by $A(k)$ and $B(k)$.  Thus the ECDE transfer function is given by
\begin{align}
\label{eq:transfer}
   T(k)\equiv \frac{\delta_{\rm DM}(k,a \gg a_\mathrm{eq})}{\delta_{\rm DM,s}(k,a \gg a_\mathrm{eq})}
    =&\frac{A(k)}{A_{\rm s}}\frac{\ln[4B(k)e^{-3}a_{\rm eq}/a_{\rm hor}(k)]}{\ln[4B_{\rm s}e^{-3}a_{\rm eq}/a_{\rm hor,s}(k)]},
\end{align}
which is independent of $a$ for $a\gg a_{\rm eq}$.  Note that Eq.~\eqref{eq:transfer} differentiates between the value of the scale factor at horizon entry in a cosmology with an ECDE ($a_{\rm hor}$) and the value of the scale factor at horizon entry in a cosmology with uninterrupted early radiation domination ($a_{\rm hor,s}$).  

Both Eqs.~\eqref{eq:log_growth} and \eqref{eq:Meszaros} ignore baryonic pressure, which prevents baryons from participating in gravitational collapse prior to their decoupling from photons and on scales smaller than their Jeans length \cite{Hu:1995en}.  For simplicity, we neglect these baryonic effects when deriving the ECDE transfer function, but we note that they can be incorporated via the procedure developed for the EMDE transfer function derived in Ref.~\cite{ES11}.  The most important consequence of baryonic pressure is that it alters $D_1(y)$ so that $\delta_{\rm DM}(a\gg a_{\mathrm{eq}}) \propto a^{0.9}$ on sub-kiloparsec scales \cite{Hu:1995en, Bertschinger06}.

We can obtain a semi-analytical estimate for the peak value of the transfer function, $T (k_{\rm pk})$ as follows. We expect the amplitude of $\delta_{\rm DM}(k_{\rm pk}, a_{\rm rh})$ at reheating to equal the amplitude of the cannibal perturbation, $\delta_{\rm c} (k_{\rm pk}, a_{\rm rh})$. 
Since the cannibal density perturbations with $k< k_J$ grow linearly after horizon entry, and the peak wavenumber $k_{\rm pk}$ enters the horizon at a scale factor $a_\mathrm{hor, pk} \propto a_{\rm fz}$, we expect $\delta_{\rm c} (k_{\rm pk}, a_{\rm rh})\propto \phi_p(k_{\rm pk}) a_{\rm rh}/a_{\rm fz}$.   This relation, particularly the assumed linear scaling of $a_\mathrm{hor, pk} =k_{\rm pk}/H(a_{\rm hor,pk}) $ with $a_{\rm fz}$, does receive logarithmic corrections owing to the cannibal interactions, but the logarithmic nature of those corrections implies that the linear scaling provides a useful estimate.  After reheating, we expect the DM perturbation to grow logarithmically.  
Thus we expect that the behavior of $\delta_{\rm DM}(k_{\rm pk})$ during the post-reheating epoch of radiation domination can be parametrized as 
 \beq
 \label{eq:dmapproxpk}
 \delta_{\rm DM} (k_{\rm pk}, a>a_{\rm rh}) = - b_1 \frac{a_{\rm rh}}{a_{\rm fz}} \phi_p (k_{\rm pk})\ln\left(b_2 \frac{a}{a_{\rm rh}}\right),
 \eeq
where $b_1$ and $b_2$ reflect, respectively, the deviation of $\delta_{\rm DM}(k_{\rm pk})$ from perfect linear growth between $a_{\rm rh}$ and $a_{\rm fz}$, and the transition of $\delta_{\rm DM}$ from linear to logarithmic growth around $a_{\rm rh}$. Empirically we find that both $b_1$ and $b_2$ are $\mathcal{O}(1)$; for instance, in the parameter point shown in Fig.~\ref{fig:cann_dom_horz_peak}, $b_1 = 2.5$ and $b_2 = 1.7$. Both $b_1$ and $b_2$ are insensitive to variations in $a_{\rm rh}/a_{\rm fz}$ as long as $\delta_{\rm DM}$ achieves linear growth by the time of reheating, or equivalently $a_{\rm rh}/a_{\rm fz}\gtrsim 200$. However, while $b_2$ is insensitive to variations in $a_{\rm fz}/a_i$, $b_1$ has a logarithmic dependence on $a_{\rm fz}/a_i$ arising from the logarithmic corrections to the assumed proportionality $a_{\rm hor,pk}\propto a_{\rm fz}$. 

\begin{figure}
	\centering
	\begin{minipage}{0.48\textwidth}
		\centering
		\includegraphics[width=1\textwidth]{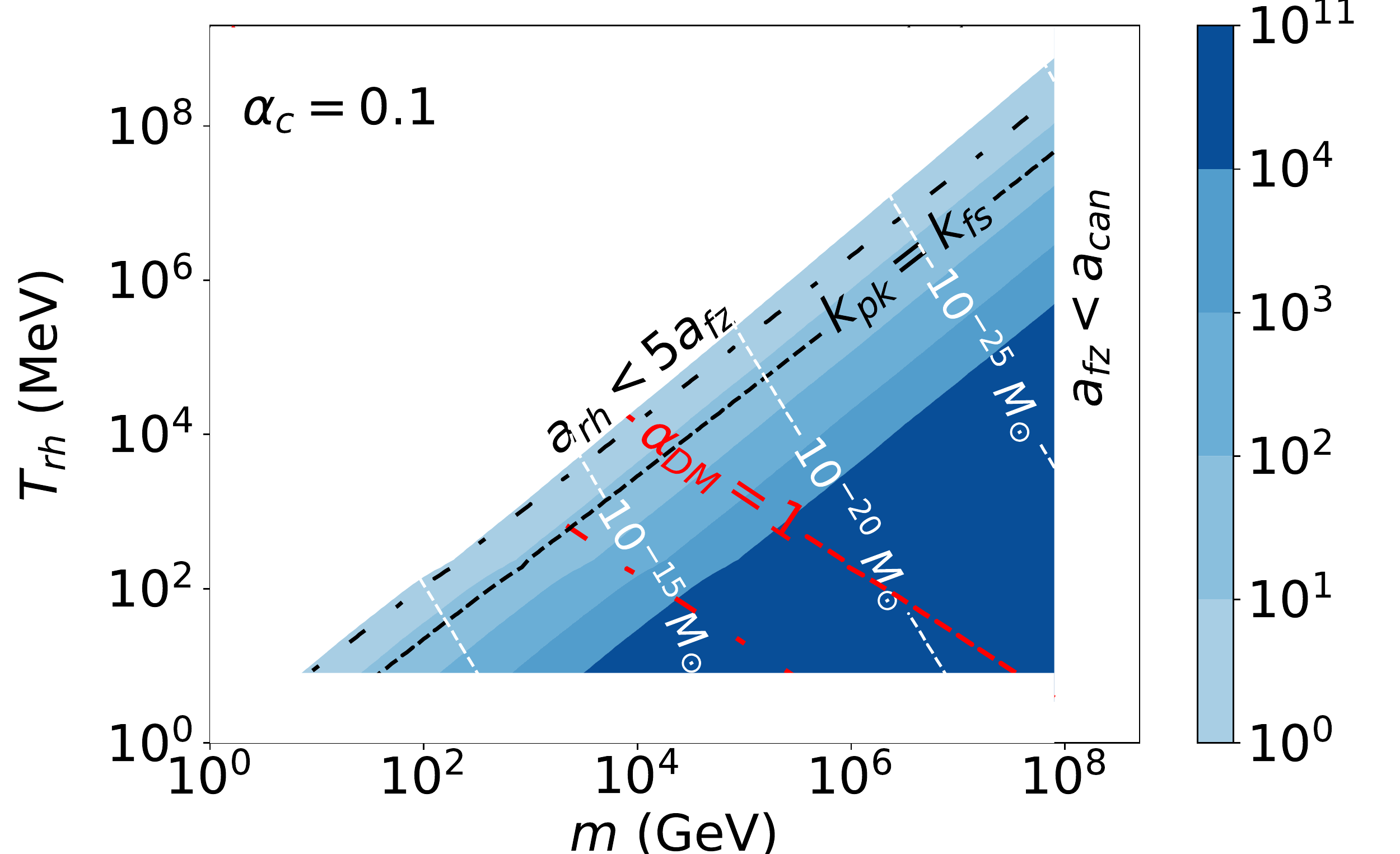}
	\end{minipage}
	\begin{minipage}{0.48\textwidth}
		\centering
		\includegraphics[width=1\textwidth]{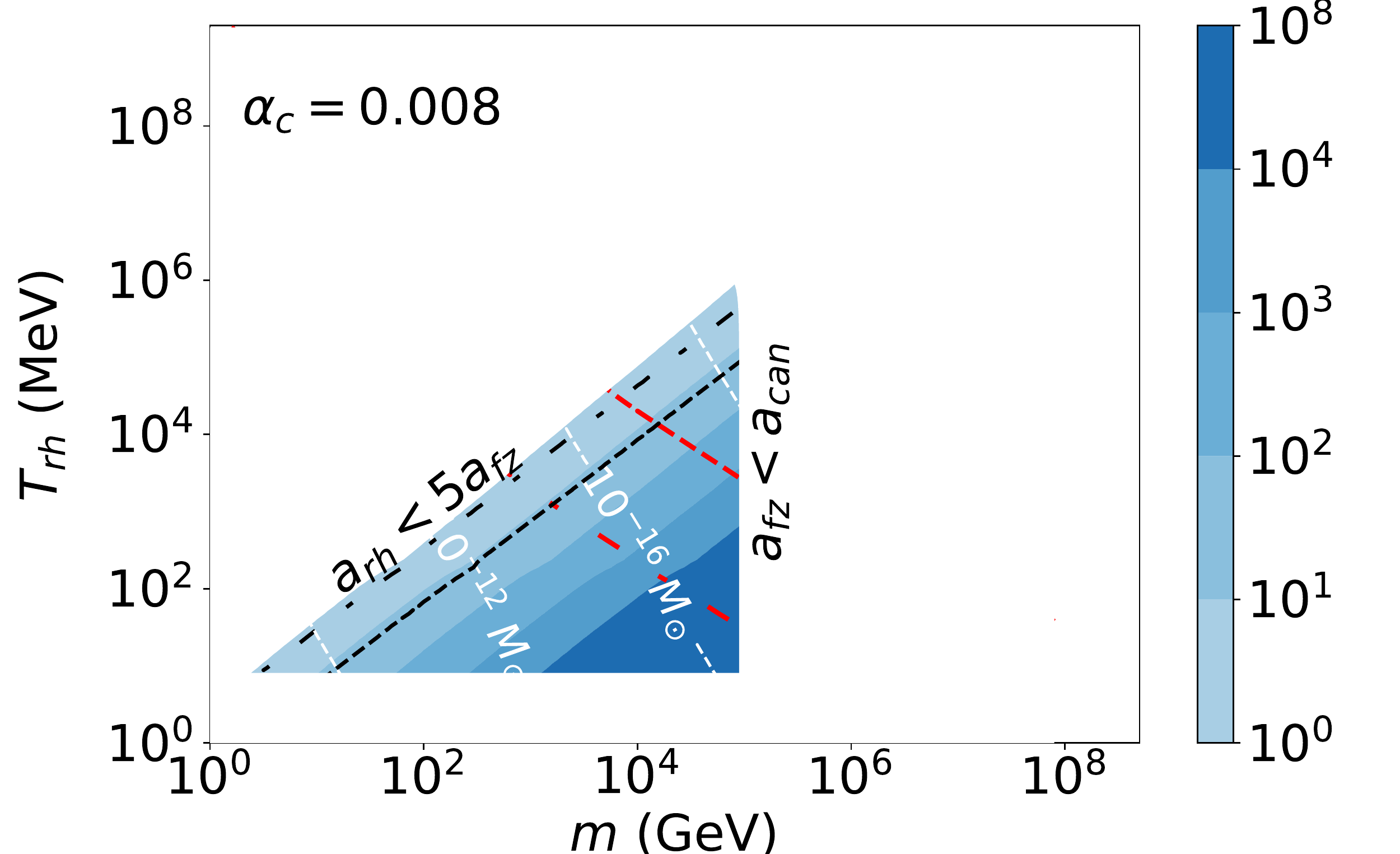}
	\end{minipage}
	\caption{Same as Fig.~\ref{fig:param_space} but for different values of $\alpha_{\rm c}$.  Smaller values of $\alpha_{\rm c}$ reduce the parameter space for cannibalism.}
\label{fig:param_space_ext}
\end{figure}

Comparing Eq.~\eqref{eq:dmapproxpk} with Eq.~\eqref{eq:log_growth} lets us identify 
 \begin{align}\label{eq:A_kpk}
A(k_{\rm pk}) = b_1\frac{a_{\rm rh}}{a_{\rm fz}} && B(k_{\rm pk})= b_2\frac{a_{\rm hor,pk}}{a_{\rm rh}}.
\end{align}
Using these relations of $A(k_{\rm pk})$ and $B(k_{\rm pk})$ in the transfer function, Eq.~\eqref{eq:transfer}, yields
\begin{align}\label{eq:pk_can}
T(k_{\rm pk})=\frac{b_1}{A_{\rm s}}\frac{a_{\rm rh}}{a_{\rm fz}}\Bigg[1-\frac{\ln(B_{\rm s}b_2^{-1}a_{\rm rh}/a_{\rm hor,s})}{\ln(4B_{\rm s}e^{-3}a_{\rm eq}/a_{\rm hor,s})}\Bigg] \sim  \frac{1}{5}\frac{a_{\rm rh}}{a_{\rm fz}},
\end{align}
where in the second relation we dropped the logarithmic factors  (since $a_{\rm rh}\ll a_{\rm eq}$) and estimated $b_1 \sim 2$.
This simple dependence of $T(k_{\rm pk})$ on ${a_{\rm rh}}/{a_{\rm fz}}$ yields the dependence of $T(k_{\rm pk})$ on $\alpha_{\rm c}, m,$ and $T_{\rm rh}$ given in Eq.~(5). Fig.~\ref{fig:param_space}, which uses a full numerical calculation of $T(k_{\rm pk})$, illustrates this dependence for $\alpha_{\rm c} = 0.02$, and we further show results for other values of $\alpha_{\rm c}$ in Fig.~\ref{fig:param_space_ext}.

\section{Dark Matter Free Streaming}
\label{app:dmfs}

The random thermal motions of DM particles will suppress perturbations on scales smaller than the DM free-streaming horizon \cite{Green_2005, Loeb_2005, Bertschinger06} and its collisionless Jeans length \cite{Piattella:2013cma}.  Since we assume that DM does not interact with the SM, the DM temperature is determined by its interactions within the hidden sector.  To quantify the DM temperature, we assume that DM kinetically decouples from the cannibal particles while nonrelativistic, which implies that the DM temperature is proportional to $a^{-2}$ after decoupling.  In the absence of gravitational forces, thermal motion will cause DM particles to traverse a comoving distance given by the free-streaming horizon:
\begin{equation}
\lambda_\mathrm{fs} = \int v_\mathrm{DM}(a) \frac{dt}{a} \approx \sqrt{\frac{T_{\rm c}(a_\mathrm{kd})}{m_\mathrm{DM}}} \int \frac{a_\mathrm{kd}}{a^3 H} da,
\label{fsdef}
\end{equation}
where $a_\mathrm{kd}$ is the value of the scale factor when DM kinetically decouples from the cannibal particles. 

In cosmologies that include an ECDE, the gravitational potential oscillates too rapidly to generate coherent gravitational forces while a perturbation mode lies within the cannibal Jeans horizon and after the decay of the cannibal gives rise to a radiation-dominated era.  While DM particles will free stream while the cannibal perturbations are oscillating, the free-streaming horizon prior to reheating is necessarily smaller than the cosmological horizon at this time.  Since the perturbation mode that maximizes the ECDE transfer function enters the horizon near the time that the cannibal freezes out, perturbations on this scale cannot be affected by DM free streaming prior to cannibal freezeout.  Moreover, even if their thermal motions cause DM particles to disperse while a given perturbation mode lies within the cannibal Jeans horizon, the gravitational forces established after that mode exits the Jeans horizon during the ECDE will cause DM to fall back into overdense regions, as illustrated by the evolution of the $k_2$ mode in the bottom left panel of Fig.~\ref{fig:cann_dom_horz_peak}.  Therefore, DM free streaming prior to reheating can only affect modes that do not exit the cannibal Jeans horizon prior to cannibal decay.  

In contrast, DM free streaming after the ECDE can suppress perturbations that would otherwise be enhanced by the ECDE.  The free-streaming horizon quickly asymptotes to a constant value after matter-radiation equality, so we can neglect dark energy when evaluating Eq.~(\ref{fsdef}).  We also make the approximations that entropy production ceases at  $a=a_\mathrm{rh}$ and $g_*(T_\mathrm{rh}) = g_{*S}(T_\mathrm{rh})$.  With our definitions of $a_\mathrm{rh}$ and $T_\mathrm{rh}$, these approximations imply that
\begin{align} 
\lambda_\mathrm{fs} \approx& \sqrt{\frac{T_{\rm c}(a_\mathrm{kd})}{m_\mathrm{DM}}} \frac{a_\mathrm{kd}}{a_\mathrm{rh}^2 \Gamma} \nonumber\\ &\times\int_{a_\mathrm{rh}}^{a_0}
  \frac{da}{a^{3}} \left[G(a) \left( \frac{1}{a}\right)^4 + G(a_\mathrm{eq})\left( \frac{1}{a^3 a_\mathrm{eq}}\right) \right]^{-1/2},
\end{align}
where $G(a)\equiv g_*(a)g_*^{1/3}(a_\mathrm{rh})/g_{*S}^{4/3}(a)$.  The quantity $\sqrt{T_{\rm c}(a_\mathrm{kd})}a_\mathrm{kd}$ increases with increasing $a_\mathrm{kd}$ up until $a_\mathrm{kd} \approx 2a_{\rm fz}$ and then becomes constant.  Therefore, the free-streaming horizon is maximized if DM decouples from the cannibal after cannibal freezeout, in which case the free-streaming horizon is independent of $a_{kd}$ because 
\begin{equation}
\sqrt{T_{\rm c}(a_\mathrm{kd})}a_\mathrm{kd} =\sqrt{T_{\rm c}(a_\mathrm{rh})}a_\mathrm{rh} = c_s(a_{\rm rh})a_\mathrm{rh}\sqrt{\frac35 {m}} .
\end{equation}
 
Perturbation modes with $k > \lambda_\mathrm{fs}^{-1}$ will be suppressed by DM free streaming \cite{Green_2005, Loeb_2005, Bertschinger06}, so the peak in the ECDE transfer function will be affected if $k_{\rm pk} \lambda_\mathrm{fs} \gtrsim 1$.  Given that $H(a_\mathrm{rh}) \approx \Gamma$, and that $k_J \propto \sqrt{a}$ after cannibal interactions freeze out,
\begin{equation}
k_{\rm pk} = k_J(2a_{\rm fz})/1.4 \approx k_J(a_{\rm rh}) \sqrt{\frac{2 a_{\rm fz}}{a_{\rm rh}}}\approx \sqrt{\frac32} \frac{a_{\rm rh} \Gamma}{c_s(a_{\rm rh})} \sqrt{\frac{2 a_{\rm fz}}{a_{\rm rh}}}.
\end{equation}
Since $T(k_{\rm pk}) \sim a_{\rm rh}/(5a_{\rm fz})$, we see that $k_{\rm pk} \lambda_\mathrm{fs}  \sim \sqrt{(9/5)m/[5 m_\mathrm{DM} T(k_{\rm pk})]}$.  Therefore, the peak scale is only affected by DM free streaming for relatively small values of $T(k_{\rm pk})$, as seen in Fig.~\ref{fig:param_space}.

In addition to DM free streaming in the absence of gravitational forces, we must also consider how the thermal motions of DM particles may prevent them from falling into the gravitational wells created by the cannibal particles on scales outside the cannibal Jeans horizon.  We have already noted that Eq.~(\ref{eq:theta_dm_eq}) for the evolution of $\theta_{\rm DM}$ neglects the anisotropic stress $\sigma$ of the dark matter particles \cite{Ma:1995ey}.  For collisionless and pressureless particles, $\sigma = -(5/3)\sigma^2_\mathrm{1D} \delta$, where $\sigma^2_\mathrm{1D} = T/m$ is the one-dimensional velocity dispersion \cite{Piattella:2013cma}: the factor of $5/3$ follows from the assumption that the particles' phase space density remains unperturbed \cite{Hogan:2000bv}.  When $\sigma$ is included in Eq.~(\ref{eq:theta_dm_eq}), Eq.~(\ref{eq:delta_DM_subhz}) becomes
\begin{widetext}
\begin{equation}
\delta_{\rm DM}''(a)+\left[\frac{(a^2 {H})'}{a^2 {H}}+\frac{1}{a}\right]\delta_{\rm DM}' +\frac{1}{a^2 }\left[\frac{ {k}^2}{(aH)^2}\left(\frac{5 T_{\rm DM}}{3m_{\rm DM}}\right)\delta_{\rm DM}-\frac{3}{2}\frac{\rho_{\rm can}}{\rho} \delta_{\rm c} \right]=0.
\label{DMjeanseqn}
\end{equation}
\end{widetext}
If the thermal motion of DM particles is neglected, $\delta_{\rm DM}$ is stagnant during the ECDE until the growth of $|\delta_{\rm c}|$ makes it equal to $\delta_{\rm DM}$, and $\delta_{\rm DM} \approx \delta_{\rm c}$ thereafter, as seen in Fig.~\ref{fig:cann_dom_horz_peak}.  To determine when thermal effects hinder the growth of $\delta_{\rm DM}$ during the ECDE, we may take $\delta_{\rm DM} \approx \delta_{\rm c}$ and $\rho_{\rm can} = \rho$ in Eq.~(\ref{DMjeanseqn}), which leads to the definition of a collisionless Jeans scale for the DM: 
\begin{equation}
k_{\rm J,DM} = aH\sqrt{\frac{9 m_{\rm DM}}{10T_\mathrm{DM}}}.
\end{equation}
If DM kinetically decouples from the cannibals prior to the end of the ECDE, modes with $k>k_{\rm J,DM}$ will be suppressed by the thermal motions of DM particles.  However, since $T_{\rm DM} \leq T_{\rm c}$ after decoupling and we assume that $m_{\rm DM} > 10 m$, $k_{\rm J,DM}$ is always larger than the cannibal Jeans wavenumber.  Therefore, the thermal motions of DM particles cannot affect the evolution of modes that exit the cannibal Jeans horizon during the ECDE, and thus do not alter the position or amplitude of the peak in the matter power spectrum generated by an ECDE.

\bibliography{cannibal}

\end{document}